\documentclass[twocolumn,showkeys,showpacs,pre]{revtex4}
\usepackage{amsfonts, amsmath, latexsym}
\usepackage[dvips]{graphicx}

\begin{document}

\title{On the efficient Monte Carlo implementation of path integrals}

\author{Cristian Predescu} 
\email{cpredescu@comcast.net}

\affiliation{
Department of Chemistry and Kenneth S. Pitzer Center for Theoretical Chemistry, University of California, Berkeley, California 94720
}
\date{\today}
\begin{abstract}
We demonstrate that the L\'evy-Ciesielski implementation of Lie-Trotter products enjoys several properties that make it extremely suitable for path-integral Monte Carlo simulations: fast computation of paths, fast Monte Carlo sampling, and the ability to use different numbers of time slices for the different degrees of freedom, commensurate with the quantum effects. It is demonstrated that a Monte Carlo simulation for which particles or small groups of variables are updated in a sequential fashion has a statistical  efficiency that is always comparable to or better than that of an all-particle or all-variable update sampler. The sequential sampler results in significant computational savings if  updating a variable costs only a fraction of the cost for updating all variables simultaneously or if the variables are independent. In the L\'evy-Ciesielski representation, the path variables are grouped in a small number of layers, with the variables from the same layer being statistically independent. The superior performance of the fast sampling algorithm  is shown to be a consequence of these  observations. Both mathematical arguments and numerical simulations are employed in order to quantify the computational advantages of the sequential sampler, the L\'evy-Ciesielski implementation of path integrals, and the fast sampling algorithm.
\end{abstract}
\pacs{02.70.Ss, 05.30.-d}
\keywords{path integrals, L\'evy-Ciesielski, random series, efficient sampling}
\maketitle

\section{Introduction}

The availability of short-time approximations having fast asymptotic convergence \cite{Tak84, Cep95, Pre04} warrants a closer look to the Monte Carlo implementation of the resulting Lie-Trotter products. The superior convergence  of path-integral methods employing such short-time approximations is achieved under the assumption that the integration against all path variables is performed in an exact fashion. In practical applications, this is never the case, except for low-dimensional problems. The efficiency of the methods suffers from the slow convergence of Monte Carlo integration.  Indeed, if the convergence order of a certain technique is $\nu$, then the computational cost to achieve a given error $\epsilon$, as measured by the number of calls to the potential function, has the form \cite{Pre04}
\begin{equation}
\label{eq:1}
\mathrm{Cost} \propto 1/\epsilon^{2+1/\nu},
\end{equation}
assuming that the Monte Carlo samples are independent. This formula demonstrates that we cannot defeat the slow convergence of the Monte Carlo simulation by indefinitely improving the convergence order.

In practical applications, Eq.~(\ref{eq:1}) represents a very optimistic evaluation, because one must deal with the additional problem of build-up of correlation among  path variables, as the number of variables increases \cite{Cep95}. Thus, only a small group of path variables can be updated in an efficient fashion at a time. Significant research has gone into the problem of diminishing the correlation between  path variables and ensuring a more efficient sampling. Techniques such as  the staging method \cite{Spr85}, the threading algorithm \cite{Pol84}, the bisection method \cite{Cep95}, the multigrid technique \cite{Jan93}, and the normal mode and Fourier approaches \cite{Dol84, Coa86, Fre86} can significantly decrease the correlation times in path integral Monte Carlo simulations. A recent technique developed in Ref.~\onlinecite{Pre04a}, which is called the fast sampling algorithm, builds upon some special properties of the so-called L\'evy-Ciesielski representation of the Feynman-Kac formula \cite{Pre02b}. The technique can be regarded as part of the random series \cite{Dol84b, Pre02} (in the continuous form) or normal mode (in the discrete form) approaches to path integration. In the present work, we demonstrate that the technique is capable of reducing the computational time necessary to achieve a prescribed statistical efficiency from $n^2$ calls to the potential function (scaling that is valid for most normal mode representations) to $n\log_2(n)$. Here, $n$ represents the number of path variables. 

In Section~II, we analyze the computational cost of the Metropolis \emph{et al} algorithm \cite{Met53, Kal86} for high dimensional systems, from the point of view of statistical efficiency. We present both mathematical and numerical arguments to justify the finding that updating particles one at a time is statistically at least as efficient as using all-particle moves. The most important cases where updating particles or path variables one at a time results in important computational savings are i) for classical systems, the case where the computational time for the whole potential increases linearly with the computational time necessary to update only one particle and ii) for path integral simulations, the case where the path variables can be grouped in independent random vectors.

In Section~III, we study the statistical efficiency of the Metropolis \emph{et al} sampler for random series as well as for the normal mode implementation of Lie-Trotter products. We conclude that the L\'evy-Ciesielski representation is superior in both cases, allowing for a reduction in the computational cost of $\log_2(n) / n$, by comparison with most normal mode implementations. The reader must realize that the fast sampling algorithm is not really a sampling technique. Rather, it is a property of the L\'evy-Ciesielski representation and can only be utilized for path integrals. Similarly, the fast computation of paths is also a property of the L\'evy-Ciesielski series, rather than a technique. It enables the computation of paths in $n\log_2(n)$ operations instead of $n^2$, the number necessary for most other normal mode implementations. It is true, starting from Coalson's Fourier-like normal mode approach \cite{Coa86}, one can still construct the paths in a time proportional to $n\log_2(n)$, by using fast sine-Fourier transform. This has been observed by Mielke and Truhlar \cite{Mie01}. However, constructing a fast sampling algorithm by using this methodology is rather difficult. Perhaps the most important property of the L\'evy-Ciesielski series is that it constitutes a link between the continuous and the discrete path integral techniques \cite{Pre02b}. Thus, almost all algorithms developed for the discrete case have an analogue in the L\'evy-Ciesielski language.  Li and Miller \cite{Li04} have recently demonstrated how a Lie-Trotter product for path integrals in many dimensions must be modified so that the number of time slices associated to each degree of freedom be proportional to the quantum effects. In the Appendix, we adapt the technique to the L\'evy-Ciesielski form and show how the number of time slices for each dimension should be chosen as a function of the particle masses.

In Section IV, we utilize the  fourth-order direct short-time approximation recently developed in Ref.~\onlinecite{Pre04}, to exemplify the use of the L\'evy-Ciesielski series for the implementation of Lie-Trotter products. We then perform Monte Carlo simulations for the $\textrm{Ne}_{19}$ Lennard-Jones cluster using both the all-variable update strategy and the fast sampling algorithm. The simulation is conducted at $4~\mathrm{K}$, using a number of $127$ path variables per degree of freedom. The numerical results demonstrate that the standard deviations for the average energy and the heat capacity estimators are more than two times larger in the case when all particles are updated simultaneously.  This translates in a computational saving of about $80\%$ if the fast sampling algorithm is utilized. However, bigger computational savings are expected for larger numbers of path variables.

\section{Considerations on the statistical efficiency of the Metropolis  sampler}

In this section, we demonstrate that the maximal displacements in the Metropolis \emph{et al} sampling algorithm decrease as fast as $n^{-1/2}$ with the number $n$ of particles that are simultaneously updated. We argue that this decrease is entropic in nature and has little to do with the interaction between particles. The generally accepted explanation for the decrease in the maximal displacements is that, by moving several of them at a time, we increase the chances that the particles collide. This explanation is mistaken and, to the contrary, we find that the decrease in the acceptance probability exists even for non-interacting particles. By means of a numerical example we show that the entropic explanation also holds for particles that interact through potentials having a strong repulsive part. 

Having quantified the decrease in statistical efficiency associated with multi-particle moves, we demonstrate that updating the particles one at a time (whether in a deterministic or random fashion) is the better strategy in terms of statistical efficiency. By statistical efficiency we understand the average distance covered by the random walker in the configuration space for a given computational time and average acceptance probability.

To begin with, let us assume that we are given a finite collection $X_1, X_2, \ldots, X_n$ of independent identically distributed random vectors (i.i.d.r.v's), taking values in some space $\mathbb{R}^d$.  These random vectors may represent, for instance, the space coordinates of a classical physical system made up of $n$ identical particles that do not interact. Let $\rho(\mathbf{x})$, with $\mathbf{x} \in \mathbb{R}^d$, be the normalized distribution of any of the random vectors $X_i$. The distribution $\rho(\mathbf{x})$ is assumed to be a smooth function, that is, to have continuous first order partial derivatives. Again, by referring to our physical system, if $V(\mathbf{x})$ is the (common) potential in which the particles move, then we may set
\[
\rho(\mathbf{x}) = e^{-\beta V(\mathbf{x})}/Q(\beta),
\]
where $\beta$ is the inverse temperature and $Q(\beta)$ is the configuration integral of the corresponding canonical system.  By independence, the overall distribution of the random vectors is given by the product $\rho(\mathbf{x}_1)\rho(\mathbf{x}_2)\ldots \rho(\mathbf{x}_n)$, which is a smooth distribution density on the space $\mathbb{R}^{dn}$.

It is perhaps clear that the best strategy for  Monte Carlo sampling of the product distribution $\rho(\mathbf{x}_1)\rho(\mathbf{x}_2)\ldots \rho(\mathbf{x}_n)$ is to perform the sampling individually, for each random vector. Thus, following Metropolis \emph{et al} \cite{Met53, Kal86}, we propose a new position for the random vector $X_i$ from the trial distribution $T(\mathbf{y}_i|\mathbf{x}_i)$, which is uniform in a $d$-dimensional hypercube centered about $\mathbf{x}_i$ and has maximal displacements $\Delta_s$, for $s = 1, 2, \ldots, d$ (therefore the sides of the hypercube have lengths $2\Delta_s$). The move is then accepted with probability \begin{equation}
\label{eq:co1}
\min\left\{1, \frac{\rho(\mathbf{y}_i)T(\mathbf{x}_i | \mathbf{y}_i)}{\rho(\mathbf{x}_i)T(\mathbf{y}_i | \mathbf{x}_i)}\right\},
\end{equation}
and rejected with the remaining probability. Repeating the procedure, one generates an ergodic Markov chain of stationary distribution $\rho(\mathbf{x}_i)$. 

Undesirable high correlation between successive positions in the Markov chain is the result of two factors i) high correlation in the proposal distribution, correlation that increases as the maximal displacements decrease, and ii) low acceptance probability. As a rule of thumb, in order to minimize the correlation, one tunes the average acceptance probability 
\begin{eqnarray}
\label{eq:co2} \nonumber
Ac_1 = \int_{\mathbb{R}^{2d}}d\mathbf{x}_id\mathbf{y}_i  \rho(\mathbf{x}_i)T(\mathbf{y}_i|\mathbf{x}_i) \\ \times
\min\left\{1,  \frac{\rho(\mathbf{y}_i)T(\mathbf{x}_i | \mathbf{y}_i)}{\rho(\mathbf{x}_i)T(\mathbf{y}_i | \mathbf{x}_i)}\right\},
\end{eqnarray}
to a value of about $50\%$, by  increasing or decreasing the maximal displacements, as appropriate \cite{Kal86}.  

 Assume now that we  sample the random vectors together and update all variables at once, using the trial distribution $T(\mathbf{y}_1|\mathbf{x}_1) \ldots T(\mathbf{y}_n|\mathbf{x}_n)$.
The move to $(\mathbf{y}_1, \mathbf{y}_2, \ldots, \mathbf{y}_n)$ is accepted  with probability
\begin{equation}
\label{eq:co3}
\min\left\{1, \prod_{i = 1}^n \frac{\rho(\mathbf{y}_i)T(\mathbf{x}_i | \mathbf{y}_i)}{\rho(\mathbf{x}_i)T(\mathbf{y}_i | \mathbf{x}_i)}\right\},
\end{equation}
and rejected with the remaining probability. The average acceptance probability is  given by the formula
\begin{eqnarray}
\label{eq:co4} \nonumber
Ac_n = \int_{\mathbb{R}^{2d}}d\mathbf{x}_1d\mathbf{y}_1 \cdots \int_{\mathbb{R}^{2d}}d\mathbf{x}_nd\mathbf{y}_n \rho(\mathbf{x}_1)T(\mathbf{y}_1|\mathbf{x}_1)\\ \cdots \rho(\mathbf{x}_n)T(\mathbf{y}_n|\mathbf{x}_n)
\min\left\{1, \prod_{i = 1}^n \frac{\rho(\mathbf{y}_i)T(\mathbf{x}_i | \mathbf{y}_i)}{\rho(\mathbf{x}_i)T(\mathbf{y}_i | \mathbf{x}_i)}\right\}. 
\end{eqnarray}

If one attempts such a strategy and utilizes the optimal maximal displacements computed for the case of  single particle moves, the average acceptance probability decreases  according to the law \cite{Pre04a}
\begin{equation}
\label{eq:co5}
Ac_n\sim e^{-Hn}, 
\end{equation}
where $H > 0$ is the relative Shannon entropy
\begin{equation}
\label{eq:co6}
H = - \int_{\mathbb{R}^{2d}}\rho(\mathbf{x})T(\mathbf{y}|\mathbf{x}) \log\left[\frac{\rho(\mathbf{y})T(\mathbf{x}| \mathbf{y})}{\rho(\mathbf{x})T(\mathbf{y} | \mathbf{x})}\right]d\mathbf{x}d\mathbf{y}.
\end{equation}
To avoid such a catastrophic decrease in the acceptance probability, we must decrease the maximal displacements, so that to minimize the Shannon entropy at a rate equal to $1/n$. More exactly, if $H_n$ is the Shannon entropy corresponding to new maximal displacements $\Delta_{s,n}$ and $Ac$ is the desired constant acceptance probability, then 
\begin{equation}
\label{eq:co7}
H_n \sim  - \log(Ac) / n.
\end{equation}

We now show that, for sufficiently large $n$, the decrease in the maximal displacements is controlled by the Fisher entropy of the smooth distribution $\rho(\mathbf{x})$. For a random variable (one-dimensional random vector), the estimate can be obtained as follows. We start with the approximation
\begin{eqnarray}
\label{eq:co8} \nonumber
H_n = - \int_{\mathbb{R}}dx\frac{\rho(x)}{2\Delta_{1,n}}\int_{-\Delta_{1,n}}^{\Delta_{1,n}} dy \log\left[{\rho(x+y)}/{\rho(x)}\right] =\\ 
- \int_{\mathbb{R}}dx\frac{\rho(x)}{2\Delta_{1,n}}\int_{-\Delta_{1,n}}^{\Delta_{1,n}} dy \log\left[1 + {\rho(x+y)}/{\rho(x)} - 1\right] \\ \approx
\int_{\mathbb{R}}dx\frac{\rho(x)}{4\Delta_{1,n}}\int_{-\Delta_{1,n}}^{\Delta_{1,n}} dy \left[{\rho(x+y)}/{\rho(x)} - 1\right]^2, \nonumber
\end{eqnarray}
where we have retained the first non-vanishing term in the logarithm expansion. This approximation becomes exact in the limit of small $\Delta_{1,n}$. In fact, in the same limit, one may expand the density $\rho(x+y)$ around the position $x$ to first order and conclude that 
\begin{eqnarray}
\label{eq:co9} \nonumber
H_n  \approx
\int_{\mathbb{R}}dx\frac{\rho(x)}{4\Delta_{1,n}}\int_{-\Delta_{1,n}}^{\Delta_{1,n}} dy \left[{\rho(x+y)}/{\rho(x)} - 1\right]^2 \\ \approx
\int_{\mathbb{R}}dx\frac{\rho(x)}{4\Delta_{1,n}}\int_{-\Delta_{1,n}}^{\Delta_{1,n}} dy \left[{\rho'(x)}/{\rho(x)}\right]^2y^2 \\ = 
\frac{\Delta_{1,n}^2}{6}\int_{\mathbb{R}} \rho'(x)^2/\rho(x)dx. \nonumber
\end{eqnarray}
The last integral appearing in the preceding formula is recognized as the Fisher entropy of the smooth distribution $\rho(x)$. For $d$-dimensional spaces, by a similar argument, the reader may obtain the general expression
\begin{eqnarray}
\label{eq:co10} 
H_n  \approx \frac{1}{6} \sum_{s = 1}^d \Delta_{s,n}^2\int_{\mathbb{R}^d} \left[{\partial_s\rho(\mathbf{x})}\right]^2/\rho(\mathbf{x})d\mathbf{x}. 
\end{eqnarray}

By comparing Eq.~(\ref{eq:co10}) with Eq.~(\ref{eq:co7}), we conclude that the asymptotic scaling of the maximal displacements in the limit of a large number of particles or random vectors that are updated simultaneously is given be the formula
\begin{equation}
\label{eq:co11}
\Delta_{s,n} \sim  \Delta_{s}^0 / \sqrt{n}.
\end{equation} 
Here, the quantities $\Delta_{s}^0$ are asymptotic constants that may have values slightly different from the optimal maximal displacements $\Delta_s$ for a one-particle or random-vector update. The decrease in the maximal displacements predicted by Eq.~(\ref{eq:co11}) is somewhat unexpected, given that the particles do not interact. In fact, the usual explanation that the decrease in the maximal displacements for multi-particle updates is the result of an increased chance in collision does not hold under closer scrutiny. As for the case of independent particles, the decrease is solely an entropic effect.

Rather than resorting to more sophisticated mathematics to demonstrate the entropic nature of the decrease in the maximal displacements, we give a numerical example, where we verify Eq.~(\ref{eq:co11}) by  performing a Monte Carlo simulation in the classical canonical ensemble for the 19-particle Lennard-Jones  cluster. We have employed the $\mathrm{Ne}_{19}$ implementation of $\mathrm{LJ}_{19}$. Although all Lennard-Jones clusters have essentially the same classical thermodynamics, as can be seen from employing reduced coordinates, we give here the exact parameters because, in the second part of the paper, we shall also use the $\mathrm{Ne}_{19}$ cluster for quantum simulations. 

The total potential energy of the $\mathrm{Ne}_{19}$ cluster is given by
\begin{equation}
\label{eq:co12}
\mathrm{V_{tot}} = \sum_{i<j}^{19} \mathrm{V_{LJ}}(r_{ij})+\sum_{i=1}^{19}
\mathrm{V_c}(\mathbf{r_i}),
\end{equation}
where $\mathrm{V_{LJ}}(r_{ij})$ is the Lennard-Jones potential describing the interaction between the particles $i$ and $j$
\begin{equation}
\label{eq:co13}
\mathrm{V_{LJ}}(r_{ij}) = 4\epsilon_{LJ}\left
        [\left( \frac{\sigma_{LJ}}{r_{ij}}\right)^{12}
       -\left( \frac{\sigma_{LJ}}{r_{ij}}\right)^{6}\right]
\end{equation} 
and $\mathrm{V_{c}}(\mathbf{r_i})$ is the confining potential
\begin{equation}
\label{eq:co14}
\mathrm{V_c}(\mathbf{r_i})=\epsilon_{LJ}\left(\frac{|\mathbf{r_i}-\mathbf{R_{cm}}|}{R_c}\right)^{20}.
\end{equation}
The role of the confining potential is to prevent the evaporation of the cluster, for the cluster by itself is not thermodynamically stable. The cluster is confined to its center of mass $\mathbf{R_{cm}}$ by a polynomial potential that increases abruptly beyond the confining radius of $R_c=2.25\sigma_{LJ}$.
The values of the Lennard-Jones parameters $\sigma_{LJ}$ and $\epsilon_{LJ}$ used are 2.749 {\AA} and 35.6 K, respectively \cite{Nei00a}. The mass of the Ne atom was set to $m_0=20.0$, the rounded atomic mass of the most abundant isotope.

The simulation has been conducted for $8$ intermediate temperatures arranged in geometric progression between $T_{min} = 0.15\epsilon_{LJ}$ and $T_{max} = 0.35\epsilon_{LJ}$. To reduce the equilibration times of the Metropolis samplers, the $8$ statistically independent parallel replicas have been involved in periodical exchanges of configurations, according to the parallel tempering algorithm \cite{Gey91, Huk96}.  For each $n = 1, 2, \ldots, 19$, the simulation has consisted of $25$ blocks of one million $n$-particle moves. The $n$ particles participating in a single move have been randomly selected from the $19$ existing particles.  After each block,  the maximal displacements are decreased or increased so that the acceptance probability for the last block is $50\%$, to a statistical accuracy  of about $0.5\%$.  

\begin{figure}[!tbp] 
   \includegraphics[angle=270,width=8.5cm,clip=t]{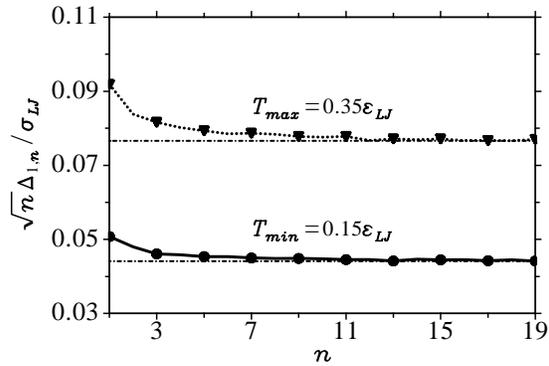} 
 \caption[sqr]
{\label{Fig:1}
Scaling of the maximal displacements with the number of particles that are simultaneously updated for the $\mathrm{LJ}_{19}$ cluster, at two different temperatures. Only every other computed values are marked on the plot. The thin lines have been added to help guide the eye toward the asymptotic region. 
}
\end{figure}

The quantities $\sqrt{n}\Delta_{1,n}$ for the replicas of lowest and largest temperatures are plotted in Fig.~\ref{Fig:1}. One can see that the asymptotic scaling predicted by Eq.~(\ref{eq:co11}) is respected to a very good degree. Similar plots for the remaining $6$ intermediate replicas show the same excellent agreement between the theoretical prediction and the  results of the simulation. It is therefore quite clear that the decrease in the maximal displacements is solely an entropic effect that has nothing to do with an increase in the chances of collision. After all, whether we move one particle at a time or all particles together, the average interactions that any group of particles suffer must be the same, at least for a well-equilibrated simulation. 

The penalty for the decrease in the maximal displacements is an increase in the correlation between successive Monte Carlo steps, increase that is due to the pronounced correlation in the proposal step. If one updates each particle individually in a deterministic or random fashion, then, on average after $n_p = 19$ moves, the position of each particle is sampled from a distribution that spans a distance proportional to $\Delta_s$. Because of the decrease in the maximal displacements, it also takes about $n_p$ Monte Carlo steps for the all-particle strategy to achieve a similar statistical efficiency, that is, to guarantee that each of the particles have been sampled from a distribution that roughly spans the same distance $\Delta_s$. 

To demonstrate the last assertion, let us look at the distances that are spanned by the random walker along some arbitrary direction, after $N$ Monte Carlo steps. Neglecting the corrections that appear because the moves are not accepted with probability one (these corrections do not change the overall scaling, as long as the acceptance probability is kept constant; in addition, not accepting the moves with probability one further reduces the distances spanned by the walker as well as the statistical efficiency of the sampler, making our final conclusion even stronger), the position of the random walker along the direction $s$ is 
\[
X_{s,N} = x_s + \Delta_{s,n_p} \sum_{k = 1}^N(2u_k - 1),
\]
where the quantities $u_k$ are independent random variables uniformly distributed on the interval $[0,1]$. For sufficiently large $N$, the sum $\sum_{k}(2u_k - 1)$ has a Gaussian distribution of variance $N / 3$ centered about the origin, as follows from the central limit theorem. Therefore, the random variable $X_{s,N}$ is a Gaussian centered about $x_s$ and of variance $\Delta_{s,n_p}^2N/3$. The average distance relative to the starting point  spanned by the random walker is 
\begin{eqnarray}
\label{eq:co15} \nonumber
\int_\mathbb{R}|z| \left(2\pi \Delta_{s,n_P}^2N/3\right)^{-1/2} e^{-z^2/\left(2\Delta_{s,n_p}^2N/3\right)}dz \\=
\sqrt{2/(3\pi)}\Delta_{s,n_p}N^{1/2} = \Delta_{s}^0\sqrt{2N/(3\pi n_p)}.
\end{eqnarray}
From Eq.~(\ref{eq:co15}), we see that it takes $N\sim n_p$ Monte Carlo steps for the all-particle strategy to achieve a statistical efficiency comparable to that of the one-particle strategy, also after $n_p$ Monte Carlo steps.

\begin{figure}[!tbp] 
   \includegraphics[angle=270,width=8.5cm,clip=t]{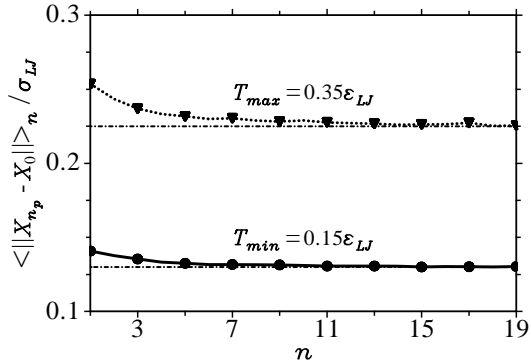} 
 \caption[sqr]
{\label{Fig:1p}
Average distances spanned by the random walker after $n_p$ Monte Carlo steps as a function of the number of particles $n$ that are updated simultaneously. The error bars are less than half the size of the plotting symbols. Only every other computed values are marked on the plot. 
}
\end{figure}

In fact, the statistical efficiency remains roughly the same no matter how many particles we move simultaneously. To illustrate this by a numerical example, we have evaluated the average distances spanned by the Monte Carlo walker after $n_p$ steps, while simultaneously updating groups of $n = 1, 2, \ldots, 19$ randomly chosen particles. If $X_N$ denotes the position of the walker at time $N$, then the average distance is 
\begin{equation}
\label{eq:co16}
\left\langle \left\|X_{n_p} - X_0 \right\| \right \rangle_n = \lim_{N\to \infty} \frac{1}{N} \sum_{k = 0}^{N - 1} \|X_{n_p + k} - X_k\|.
\end{equation}
In collecting the averages, one must discard all differences $\|X_{n_p + k} - X_k\|$ for which a parallel tempering swap has occurred at any Monte Carlo step between $k$ and $k + n_p$. The average distances are shown in Fig.~\ref{Fig:1p} and are seen to closely mimic the behavior of the maximal displacements. Thus, the one-particle and the multiparticle updating strategies have essentially the same statistical efficiency. 

In practical applications, updating the particles one at a time is almost always the winning strategy in terms of computational effort for same statistical efficiency.  In many applications,  the potential can be decomposed in $n_p$ smaller parts, each describing the interaction of a particle with its environment, and each taking $n_p$-times less computational effort to evaluate. Thus, the single-particle strategy ensures that each particle is sampled from a distribution spanning a distance of $\Delta_s$, in a time roughly equal to the time for a single all-particle update. The all-particle strategy takes $n_p$-times more computational resources to achieve similar results. But even for the situations where such a decomposition is not possible, the sequential sampling is superior because it allows for a better tuning of the maximal displacements. In the all-particle strategy, the optimal ratios between the various maximal displacements cannot be determined during the simulation and have to be fixed a priori. 

Let us conclude this section by mentioning that the decrease in the maximal displacement of $n^{-1/2}$ is solely due to the larger number of particles that are updated simultaneously. If the strength of the correlation increases with the number of particles, then the decrease in the maximal displacement for the multiparticle update is $n^{-1/2}$ times the decrease in the maximal displacement for the one-particle update. For instance, consider the problem of sampling the distribution
\begin{equation}
\label{eq:co17}
\exp\left\{-\frac{1}{2} \frac{ (x_0 - x_1)^2 + (x_1 - x_2)^2 + \ldots + (x_n - x_1)^2} {\sigma^2 / n}\right\},
\end{equation}
where $\sigma^2 = \hbar^2\beta / m_0$. This distribution is encountered in the construction of Lie-Trotter products. The decrease in the maximal displacement for the all-particle update strategy is $n^{-1/2} \times n^{-1/2}$, with the first factor due to the larger number of particles and the second factor due to the decrease in the maximal displacement for one-particle moves. Thus, the number of Monte Carlo steps necessary to achieve a prescribed statistical efficiency is $n^2$. Since the computational effort for a single Monte Carlo step in terms of calls to the potential is also proportional to $n$, we see that the cost for the direct Monte Carlo sampling of the Trotter-Lie products is proportional to $n^3$, result consistent with the one obtained in Ref.~\cite{Cep95}.

\section{Statistical efficiency for path integral sampling}

In the preceding section, we have demonstrated that the efficiency of the sequential sampler cannot be defeated by employing all-particle moves. This finding simplifies the efficiency study for the different path-integral sampling strategies. In this section, we shall analyze the statistical efficiency of the random series approach to path integrals and of related implementations. We shall see that the computational time for a given statistical efficiency scales as $n$ times the cost to evaluate the action, for most series. Here, $n$ is the number of path variables. One important exception is the L\'evy-Ciesielski series, for which the scaling is $\log_2(n)$ times the cost to evaluate the action. This remarkable property of the L\'evy-Ciesielski series constitutes the engine behind the fast sampling algorithm \cite{Pre04a}. 

In the second subsection, we specialize the findings obtained in the case of random series for the normal mode implementation of Lie-Trotter products. For such products, the time to evaluate the action is proportional to the number of path variables. Therefore, the computational time for a given statistical efficiency is proportional to $n^2$ for most normal mode approaches, except for the L\'evy-Ciesielski one, for which the scaling is $n \log_2(n)$. 

\subsection{Random series implementation of path integrals}
A standard approach for the numerical implementation of the Feynman-Kac formula \cite{Fey48, Kac51, Sim79} is via  random series \cite{Pre02, Dol84b}. The implementation is as follows. Let $\{\lambda_k(\tau)\}_{k \geq 0}$ be any orthonormal basis in $L^2[0,1]$ such that $\lambda_0(\tau) = 1$. Define the primitives
\begin{equation}
\label{eq:st1}
\Lambda_k(u)=\int_{0}^{u}\lambda_k(\tau)d\tau.
\end{equation}
Let $\Omega$ denote the set of all sequences $\bar{a} := \{a_1, a_2, \ldots\}$. The
Gaussian measure 
\begin{equation}
\label{eq:st2}
dP[\bar{a}] = \prod_{k = 1}^\infty \frac{1}{\sqrt{2\pi}}e^{-a_k^2/2}da_k
\end{equation}
on $\Omega$ makes the normal random variables $\bar{a} := \{a_1, a_2, \ldots\}$ independent and identically distributed. With the above notations, the one-dimensional Feynman-Kac formula reads \cite{Pre02} 
\begin{eqnarray}
\label{eq:st3}\nonumber&&
{\rho(x,x';\beta)}=\rho_{fp}(x,x';\beta)\int_{\Omega}dP[\bar{a}] \\ && \times  \exp\left\{-\beta\int_{0}^{1}\! \!  V\left[x_r(u)+\sigma \sum_{k = 1}^\infty a_k \Lambda_k(u)\right] d u\right\}. \qquad
\end{eqnarray}
Eq.~(\ref{eq:st3}) is called the random series representation of the Feynman-Kac formula. The quantities $\rho(x,x';\beta)$ and $\rho_{fp}(x,x';\beta)$ represent the density matrices of the physical system and of the free particle, respectively. $x_r(u)$ stands for $x + (x'-x)u$, whereas $\sigma = (\hbar^2 \beta /m_0)^{1/2}$. The generalization to many dimensions is straightforward: one just considers an independent random series for each additional physical degree of freedom. 

The random series representation of the Feynman-Kac formula is made possible by the Ito-Nisio theorem \cite{Pre02, Kwa92}, which gives an explicit construction of the Brownian bridge entering the Feynman-Kac formula. This theorem implies that the Feynman-Kac formula is invariant to orthonormal transformations corresponding to changes from a basis $\{\lambda_{k}(u)\}_{k \geq 1}$ orthogonal on $\lambda_0 = 1$ to another basis $\{\lambda'_k(u)\}_{k\geq 1}$, also orthogonal on the constant function. The reader may easily rationalize this observation by noticing that the measure defined by Eq.~(\ref{eq:st2}) is invariant under an orthonormal transformation $a'_k = \sum_{j \geq 1} \theta_{k,j}a_j$. 

Important examples of  series representations of the Feynman-Kac formula are provided by the Wiener-Fourier series and  L\'evy-Ciesielski series. The Wiener-Fourier series representation is obtained from the cosine Fourier basis $\{\lambda_k(\tau) = \sqrt{2}\cos(k \pi \tau)\}_{k \geq 1}$, which, together with $\lambda_0(\tau) = 1$, forms a complete orthonormal basis of $L^2[0,1]$. The primitives of the cosine functions are
\[
\Lambda_{k}(u)= \int_0^u \lambda_k(\tau)d\tau = \sqrt{\frac{2}{\pi^2}}\frac{\sin(k\pi u)}{k}.
\]
Upon replacement in Eq.~(\ref{eq:st3}), we obtain
\begin{eqnarray}
\label{eq:st4}\nonumber&&
{\rho(x,x';\beta)}=\rho_{fp}(x,x';\beta)\int_{\Omega}dP[\bar{a}]  \exp\Bigg\{-\beta \\ && \times \int_{0}^{1}\! \!  V\left[x_r(u)+\sigma \sum_{k = 1}^\infty a_k \sqrt{\frac{2}{\pi^2}}\frac{\sin(k\pi u)}{k} \right] d u\Bigg\}. \qquad
\end{eqnarray}
Eq.~(\ref{eq:st4}) has been first utilized in the context of path integrals by Doll and Freeman \cite{Dol84b}. As an application that has no analogue for discrete path integral techniques, they  have observed that the random series representation enables a computational technique, called partial averaging \cite{Dol85}, that has been recently shown to converge for all physically reasonable potentials \cite{Pre03}. 

A second important random series representation is the L\'evy-Ciesielski one. In this case, one starts with the so-called Haar basis, which is made up of the functions
\begin{equation}
\label{eq:st5}
f_{k,j}(\tau)=\left\{\begin{array}{cc} 2^{(k-1)/2},& \tau \in [(l-1)/2^k, l/2^k]\\ - 2^{(k-1)/2},& \tau \in [l/2^k, (l+1)/2^k]\\ 0, &\text{elsewhere,} \end{array}\right.
\end{equation}
where $l=2j-1$.
Together with $f_0\equiv 1$, these functions make up a complete orthonormal basis in $L^2([0,1])$. Their primitives 
\begin{widetext}
\begin{equation}
\label{eq:st6}
F_{k,j}(u)=\left\{\begin{array}{cc} 2^{(k-1)/2}[u-(l-1)/2^k],& u \in [(l-1)/2^k, l/2^k]\\ 2^{(k-1)/2}[(l+1)/2^k-u],& u \in [l/2^k, (l+1)/2^k]\\ 0, &\text{elsewhere} \end{array}\right.
\end{equation}
\end{widetext}
are called the \emph{Schauder functions}. As McKean puts it \cite{McK69}, the Schauder functions are ``little tents,'' which can be obtained one from the other by dilatations and translations. In modern terminology, this has to do with the fact that the original Haar wavelet basis is a multiresolution analysis of $L^2([0,1])$  organized in ``layers'' indexed by $k$ \cite{Mal99}.  

For $k=1,2,\ldots$ and $j=1,2,\ldots,2^{k-1}$, the Schauder functions $F_{k,j}(u)$ are generated by translations and dilatations of the function 
\begin{equation}
\label{eq:st7}
F_{1,1}(u) = \left\{\begin{array}{cc} u,& u \in (0, 1/2],\\ 1-u,& u \in (1/2, 1),\\ 0, &\text{elsewhere}. \end{array}\right.
\end{equation}
More precisely, we have
\begin{equation}
\label{eq:st8}
F_{k,j}(u)= 2^{-(k-1)/2} F_{1,1}(2^{k-1}u - j + 1),
\end{equation}
for  $k \geq 1$ and $1 \leq j \leq 2^{k-1}$.

\begin{figure}[!tbp] 
   \includegraphics[angle=270,width=8.5cm,clip=t]{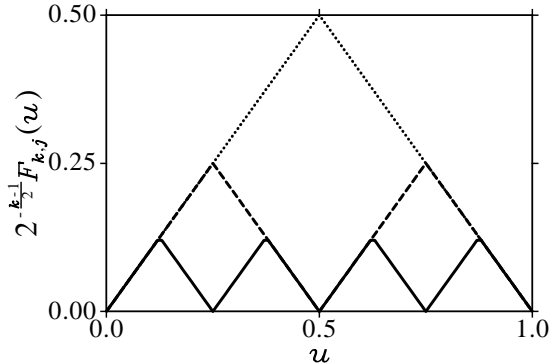} 
 \caption[sqr]
{\label{Fig:2}
A plot of the renormalized Schauder functions for the layers $k=1,2,\,\text{and}\,3$, showing the pyramidal structure.
}
\end{figure}

If we multiply them by $2^{-(k-1)/2}$, the Schauder functions make up a pyramidal structure organized in layers indexed by $k$, as shown in Fig.~\ref{Fig:2}.
The supports (the sets on which the functions do not vanish) of the Schauder functions are the open intervals of the form $(u_{k,j-1}, u_{k,j})$, for $1 \leq j \leq 2^{k-1}$, where $u_{k,j} = j2^{-(k-1)}$. The supports are \emph{disjoint} for functions corresponding to the same layer $k$. Because of this property, we have the equality 
\begin{equation}
\label{eq:st9}
\sum_{j = 1}^{2^{k-1}}a_{k,j} F_{k,j}(u) = a_{k, [2^{k-1}u]+1}F_{k,[2^{k-1}u] + 1}(u),
\end{equation}
for any sequence of numbers $a_{k,1}, a_{k,2}, \ldots , a_{k,2^{k-1}}$. Here, $[x]$ denotes the largest integer smaller or equal to $x$, whereas for $u = 1$, the quantities $a_{k, 2^{k-1} + 1}$ and $F_{k, 2^{k-1}+1}(1)$ are defined to be equal to $0$.

In the new representation, the Feynman-Kac formula reads \cite{Pre02b}
\begin{widetext}
\begin{eqnarray}
\label{eq:st10}&&
{\rho(x,x';\beta)}=\rho_{fp}(x,x';\beta)\int_{\Omega}dP[\bar{a}]  \exp\left\{-\beta  \int_{0}^{1}\! \!  V\left[x_r(u)+\sigma \sum_{l = 1}^\infty a_{l, [2^{l-1}u]+1}F_{l,[2^{l-1}u] + 1}(u) \right] d u\right\}. 
\end{eqnarray}
\end{widetext}
In the L\'evy-Ciesielski representation, the independent random variables $a_1, a_2, \ldots$ have been re-indexed as $\{a_{l,j}; l=1,2,\ldots; j=1,2,\ldots,2^{l-1}\}$, in agreement with the indexing scheme employed for the series. 

The numerical advantages of the L\'evy-Ciesielski representation are multiple. Assume that we truncate the series up to a number of $n = 2^k - 1$ path variables. That is, we use exactly $k$ complete layers. Given $u \in [0,1]$, we only need $k = \log_2(n+1)$ operations to perform the evaluation of the series at the point $u$. This is in contrast with the Wiener-Fourier series, for which one needs $n$ operations. This property is called fast computation of paths \cite{Pre02b}. 

A second property, which is called the fast sampling property \cite{Pre04a},  has to do with the sampling of the paths. We have already demonstrated in the preceding section that the efficiency of the sequential sampling technique cannot be exceeded by the techniques employing multi-variable updates.  Also, notice that the maximal displacements for the individual update of the different path variables, although not equal, do not decrease to zero. For path variables of large indexes, they converge to the maximal displacements for a normally distributed random variable.  (This observation is also true for the normal mode representation of Lie-Trotter products, considered in the following section).  Since a complete sweep through the space of path variables is done in $n$ steps,  it follows that the computational effort to achieve a prescribed statistical efficiency for the Wiener-Fourier series is $n$ times the cost to evaluate the action (we shall call action the one-dimensional integral over the interval $[0,1]$ appearing at the exponent). This cost does not change if an all-variable sampling strategy is adopted. 

For the L\'evy-Ciesielski series, one still needs to update each path variable one at a time. However, it is not necessary to compute the whole action in order to update a variable. More precisely, if the variable $a_{l,j}$ is to be updated, then one only needs to compute the quantity
\begin{widetext}
\begin{eqnarray}
\label{eq:st11}&&
e^{-a_{l,j}^2} \exp\left\{-\beta  \int_{(j-1)2^{-(l-1)}}^{j2^{-(l-1)}}\! \!  V\left[x_r(u)+\sigma \sum_{l = 1}^k a_{l, [2^{l-1}u]+1}F_{l,[2^{l-1}u] + 1}(u) \right] d u\right\}, 
\end{eqnarray}
\end{widetext}
in order to make the decision if the variable $a_{l,j}$ is to be updated or not, according to the Metropolis \emph{et al} criterion. 
The point here is that the terms
\begin{eqnarray*} 
&&
\exp\bigg\{-\beta  \int_{0}^{(j-1)2^{-(l-1)}}\! \!  V\bigg[x_r(u)+\sigma \\ && \times \sum_{l = 1}^k a_{l, [2^{l-1}u]+1}F_{l,[2^{l-1}u] + 1}(u) \bigg] d u\bigg\}. 
\end{eqnarray*}
and
\begin{eqnarray*}
&&
\exp\bigg\{-\beta  \int_{j2^{-(l-1)}}^{1}\! \!  V\bigg[x_r(u)+\sigma \\ && \times \sum_{l = 1}^k a_{l, [2^{l-1}u]+1}F_{l,[2^{l-1}u] + 1}(u) \bigg] d u\bigg\}. 
\end{eqnarray*}
do not contain the variable $a_{l,j}$, because the functions $F_{l,j}(u)$ vanish outside the open interval
\[\left((j-1)2^{-(l-1)}, j2^{-(l-1)}\right).\] Therefore, with a single evaluation of the action, we can update all $2^{l-1}$ variables from the layer $l$, independently. Thus, the computational cost is the product between the number of layers $k = \log_2(n+1)$ and the cost to evaluate the action. The L\'evy-Ciesielski representation is $n / \log_2(n+1)$ times faster than the Wiener-Fourier series from the point of view of sampling efficiency.

\subsection{Sampling efficiency for the normal mode approach to Lie-Trotter products}

The traditional way of constructing approximations to the Feynman-Kac formula is via Lie-Trotter products \cite{Tro59, Fey65, Cep95}. Such a construction starts with a short-time high-temperature approximation to the density matrix, say $\rho_0(x,x';\beta)$. Because the density matrix of a free particle is strictly positive, any such short-time approximation  can be put in the product form
\begin{equation}
\label{eq:st12}
\rho_0(x,x';\beta) = \rho_{fp}(x,x';\beta)r_0(x,x';\beta). 
\end{equation}
Letting $x_0 = x$, $x_{n+1} = x'$, and $u_i = i /(n+1)$ for $0 \leq i \leq n+1$,  the $n$-th order  Lie-Trotter product obtained from the short-time approximation considered above takes the form
\begin{eqnarray}
\label{eq:st13}
\nonumber
\rho_n(x,x';\beta) =\int_{\mathbb{R}^n}  \prod_{i = 0}^{n}p_{\sigma^2(u_i - u_{i+1})}(x_i,x_{i+1}) \\ \times \prod_{j = 0}^n r_0(x_{j},x_{j+1};\beta/2^k) dx_1 \cdots dx_n. 
\end{eqnarray}
Here,  $p_u(x,x')$ is defined by 
\begin{equation}
\label{eq:st14}
p_u(x,x') = (2\pi u)^{-1/2}\exp\left[-{(x'-x)^2}/({2u})\right].
\end{equation}

A set of $n$ Gaussian random variables having joint probability distribution
\begin{equation}
\label{eq:st15}
\prod_{i = 0}^{n}p_{\sigma^2(u_i - u_{i+1})}(x_i,x_{i+1})dx_1\ldots dx_n
\end{equation}
can be constructed in various ways \cite{But55, Coa86, Pre02b}. For instance, by diagonalization, Coalson \cite{Coa86}  has shown that if $a_1, a_2, \ldots, a_n$ are independent Gaussian variables of mean zero and variances 
\begin{equation}
\label{eq:st16}
\lambda_i=4\sigma^2(n+1)\sin^2\left[\frac{i\pi}{2(n+1)}\right],\quad 1\leq i \leq n,
\end{equation}
then 
\begin{equation}
\label{eq:st17}
x_r(u_i) + \sum_{j = 1}^n a_j S_{i,j}
\end{equation}
has the distribution given by Eq.~(\ref{eq:st15}). Here, 
\begin{equation}
\label{eq:st18}
S_{i,j}=\sqrt{\frac{2}{n+1}}\sin\left(\frac{ij\pi}{n+1}\right),\quad 1\leq i, j \leq n.
\end{equation}

As argued in Refs.~\onlinecite{Pre02, Pre02b}, a more useful form is 
\begin{equation}
\label{eq:st19}
x_r(u_i) + \sigma \sum_{j = 1}^n a_j  S_{i,j}/\lambda_j^{1/2}
\end{equation}
with the variables $a_1, a_2, \ldots, a_n$ being independent identically distributed normal random variables. The main reason is that the temperature dependence is now buried into the expression of the short-time approximation. In the limit of large $n$, the resulting discrete approximation 
\begin{widetext}
\begin{eqnarray}
\nonumber 
\label{eq:st20}
\rho_n(x,x';\beta) &=&\rho_{fp}(x,x';\beta) \int_{\mathbb{R}}da_{1} \cdots \int_{\mathbb{R}}da_{n} (2\pi)^{-n/2}\prod_{k = 1}^{n}e^{-a_{k}^2/2}  \\ &&\times \prod_{i = 0}^{n} r_0\left[x_r(u_i)+\sigma \sum_{j=1}^n S_{i,j}{a_j}/{\lambda_j^{1/2}}, x_r(u_{i+1})+\sigma \sum_{j=1}^n S_{i+1,j}{a_j}/{\lambda_j^{1/2}} ;\frac{\beta}{n+1}\right]. 
\end{eqnarray}
\end{widetext}
converges to the Feynman-Kac formula in re-scaled form. Therefore, the thermodynamic estimators obtained from formal differentiation against the inverse temperature  have finite variance in the limit of large number of path variables \cite{Pre02, Pre03b, Pre03c}. 

As for random series, Eq.~(\ref{eq:st20}) is invariant under orthogonal transformations. In the present form, the formula looks like the Wiener-Fourier series. However, as argued in Ref.~\onlinecite{Pre02b}, by appropriate orthogonal transformations, the representation given by Eq.~(\ref{eq:st20}) can be made to look like any series we want [more precisely, in the limit of large $n$, we can make Eq.~(\ref{eq:st20}) look like any series allowed by the Ito-Nisio theorem]. If $n = 2^k - 1$, another possible construction of a set of $n$ Gaussian variables having the joint distribution given by Eq.~(\ref{eq:st15}) is 
\begin{equation}
\label{eq:st21}
x_r(u_j) + \sigma  \sum_{l=1}^{k}a_{l, [2^{l-1}u_{j}]+1}F_{l,[2^{l-1}u_{j}] + 1}(u_{j}).
\end{equation}
The exact orthogonal transformation that takes Eq.~(\ref{eq:st19}) into Eq.~(\ref{eq:st21}) does not really matter, as Eq.~(\ref{eq:st21}) can be demonstrated directly from the L\'evy-Ciesielski representation of the Brownian bridge \cite{Pre02b, Pre04a}. 
The Lie-Trotter product now becomes
\begin{widetext}
\begin{eqnarray}
\nonumber 
\label{eq:st22}
\rho_n(x,x';\beta) &=&\rho_{fp}(x,x';\beta) \int_{\mathbb{R}}da_{1,1} \cdots \int_{\mathbb{R}}da_{k,2^{k-1}} (2\pi)^{-n/2}\prod_{l = 1}^{k}\prod_{i = 1}^{2^{l-1}}\exp\left(-a_{l,i}^2/2\right)  \\ &&\times \prod_{j = 0}^{n} r_0\left[x_r(u_j) + \sigma \sum_{l=1}^{k}a_{l, [2^{l-1}u_{j}]+1}F_{l,[2^{l-1}u_{j}] + 1}(u_{j}), \right.\\&& \left.  x_r(u_{j+1}) + \sigma \sum_{l=1}^{k}a_{l, [2^{l-1}u_{j+1}]+1}F_{l,[2^{l-1}u_{j+1}] + 1}(u_{j+1});\beta/2^k\right]. \nonumber
\end{eqnarray}
\end{widetext}

Eq.~(\ref{eq:st22}) has the same numerical advantages over Coalson's sine-Fourier form as the L\'evy-Ciesielski representation has over the Wiener-Fourier series: fast computation and  sampling of paths. The analysis performed in the preceding section carries over here in a simple form. The computation and sampling of paths is done in $(n+1) \log_2(n+1)$ operations for the L\'evy-Ciesielski representation because there are $n + 1$  slices. More precisely, for sampling, $(n +1) \log_2(n+1)$ represents the number of calls to $r_0[x,x';\beta / 2^k]$ in order to update all path variables sequentially. For Coalson's sine-Fourier form, one needs $(n+1)^2$ operations two perform the sampling (whether sequential or all-variable updates are attempted) in order to ensure a given statistical efficiency of the sampler. The computation of paths also take $(n+1)^2$ operations if implemented directly. However, the computation of paths can be done in $(n+1)\log_2(n+1)$ operations by means of the fast sine-Fourier transform, as pointed out by Mielke and Truhlar \cite{Mie01}. Most likely, the orthogonal transformation that takes Eq.~(\ref{eq:st19}) into Eq.~(\ref{eq:st21}) is the one that enables the sine-Fourier transform algorithm. 

In the L\'evy-Ciesielski representation, the quantity that must be used to test if the variable $a_{l,i}$ is updated or not is
\begin{widetext}
\begin{eqnarray}
\nonumber 
\label{eq:st23}
e^{-a_{l,i}^2/2}  \prod_{j = (i-1)2^{k-l+1} }^{i2^{k-l+1} - 1 } r_0\left[x_r(u_j) + \sigma \sum_{l=1}^{k}a_{l, [2^{l-1}u_{j}]+1}F_{l,[2^{l-1}u_{j}] + 1}(u_{j}),  \right.\\  \left. x_r(u_{j+1}) + \sigma \sum_{l=1}^{k}a_{l, [2^{l-1}u_{j+1}]+1}F_{l,[2^{l-1}u_{j+1}] + 1}(u_{j+1});\beta/2^k\right]. 
\end{eqnarray}
\end{widetext}
If the short-time approximation introduces additional path variables, then these path variables are to be sampled separately. They can be grouped into $2^k$ independent subsets, which can be individually tested for acceptance. We shall give such an example in the following section. 

Let us address the issue of overall efficiency for the Monte Carlo simulation of Lie-Trotter products. Assume the order of convergence of the short-time approximation is $\nu$. To achieve a final error of $\epsilon$, we need to utilize $n \propto \epsilon^{-1/\nu}$ path variables. The computational cost to efficiently update all path variables once is proportional to $n^2 \propto \epsilon^{-2/\nu}$ for the Wiener-Fourier approach and to $n \log_2(n) \propto \epsilon^{-1/\nu}\nu^{-1}\log_2(1/\epsilon)$ for the L\'evy-Ciesielski form. This cost must be multiplied by the number of steps necessary for the Monte Carlo sampler  to reach an error of $\epsilon$, number of steps that is proportional to $\epsilon^{-2}$. Thus, the overall cost is 
\begin{equation}
\label{eq:st24}
\textrm{Cost} \propto \epsilon^{-2-2/\nu},
\end{equation}
for the Wiener-Fourier approach, and 
\begin{equation}
\label{eq:st25}
\textrm{Cost} \propto \nu^{-1}\epsilon^{-2-1/\nu}\log_2(1/\epsilon),
\end{equation}
for the L\'evy-Ciesielski approach, respectively. The scaling for the L\'evy-Ciesielski approach is only marginally worse than the ideal scaling expressed by Eq.~(\ref{eq:1}).

\section{An application of the L\'evy-Ciesielski implementation}

In this section, we illustrate the numerical advantages of the L\'evy-Ciesielski implementation and the fast sampling algorithm by utilizing the technique in the context of the direct fourth-order short-time approximation introduced in Ref.~\onlinecite{Pre04}. The resulting path-integral expressions are then employed to compute the energy and the heat capacity of the $\mathrm{Ne}_{19}$ cluster, at the temperature of $4~\mathrm{K}$. We perform two simulations using the fast sampling algorithm and the all-variable sampling strategy. Important reductions in the statistical errors of the thermodynamic energy and heat capacity estimators are observed for the fast sampling algorithm. These reductions are solely due to the decrease in correlation between the successive steps of the generated Monte Carlo Markov chain.

The fourth-order short-time approximation is given by the formula
\begin{eqnarray}
\label{eq:ap1} \nonumber &&
r_0(x,x';\beta) = \int_{\mathbb{R}^3}(2\pi)^{-3/2} e^{-(b_1^2 + b_2^2 + b_3^2)/2}\\ && \times \exp\left\{-\beta \sum_{k = 1}^4 \omega_k V\left[x_r(\theta_k) + \sigma \sum_{j = 1}^3 b_j \tilde{\Lambda}_j(\theta_k)\right]\right\}. \qquad
\end{eqnarray}
In Eq.~(\ref{eq:ap1}), $\omega_k$ and $\theta_k$ are the weights and points for the four-point Gauss-Legendre quadrature technique on the interval $[0,1]$. They are given in Table~\ref{Tab:I} for ease of reference.
\begingroup
\begin{table}[!tbp]
\caption{
\label{Tab:I}
Quadrature points and weights for the $4$-point Gauss-Legendre technique on the interval $[0,1]$. }
\begin{tabular}{|c |c |c |c |c |}
\hline
$i$ & 1 & 2 & 3 & 4 \\ 
\hline \hline
$\theta_i$&0.069431844&0.330009478&0.669990522&0.930568156\\
\hline
$\omega_i$&0.173927423&0.326072577&0.326072577&0.173927423\\
\hline 
\end{tabular}
\end{table}
\endgroup
The three functions $\tilde{\Lambda}_j(u)$ are defined by the equations
\begin{eqnarray}
\label{eq:ap2}
\left\{\begin{array}{ll} 
\tilde{\Lambda}_1(u) =&  \sqrt{3}u(1-u),\\
\tilde{\Lambda}_2(u) =& r(u) \cos[\alpha_1(u-0.5)+ \alpha_2(u-0.5)^3], \\ 
 \tilde{\Lambda}_3(u)=& r(u) \sin[\alpha_1(u-0.5)+ \alpha_2(u-0.5)^3], 
\end{array}\right.
\end{eqnarray}
with
\[
r(u)= \left\{u(1-u)[1-3u(1-u)]\right\}^{1/2}.
\]
The numerical values of the constants $\alpha_1$ and $\alpha_2$ are
\begin{equation}
\label{eq:ap3}
\alpha_1 \approx 6.379716466 \quad \text{and} \quad \alpha_2 \approx 8.160188248.
\end{equation}

Using Eq.~(\ref{eq:st22}), we can arrange the additional path variables as supplementary layers in the L\'evy-Ciesielski series. Extend the functions $\{\tilde{\Lambda}_l(u); 1 \leq l \leq 3\}$ outside the interval $[0,1]$ by setting them to zero  and define
 \begin{equation}
\label{eq:ap4}
G^{(l)}_{k,j}(u) = 2^{-k/2}\tilde{\Lambda}_l(2^k u - j + 1),
\end{equation}
for $1\leq l \leq 3$ and $1 \leq j \leq 2^k$.  Then, with the convention that $a_{l, 2^{l-1}+1} = 0$, for $1 \leq l\leq k$, and $b_{l,2^k+1} = 0$, for $l = 1, 2, 3$, we have \cite{Pre04}
\begin{widetext}
\begin{eqnarray}
\label{eq:ap5} \nonumber
\frac{\rho_n(x, x' ;\beta)}{\rho_{fp}(x, x' 
;\beta)}&=&\int_{\mathbb{R}} d a_{1,1}\ldots \int_{\mathbb{R}} d a_{k,2^{k-1}}  \left( 2\pi \right)^{-n/2}  \exp\left({-\frac{1}{2}\sum_{l=1}^{k}\sum_{j=1}^{2^{l-1}}  a_{l,j}^2}\right) \\ & \times & \nonumber
\int_{\mathbb{R}} d b_{1,1}\ldots \int_{\mathbb{R}} d b_{3,2^k}  \left( 2\pi \right)^{-3(n+1)/2}  \exp\left({-\frac{1}{2}\sum_{l=1}^{3}\sum_{j=1}^{2^{k}}  b_{l,j}^2}\right)
\\& \times & \exp\left\{-\beta \int_0^1 V\left[x_r(u)+\sigma \sum_{l=1}^{k} a_{l,[2^{l-1} u]+1} \;{F}_{l,[2^{l-1} u]+1}(u) \right. \right. \\ && \left. \left.+ \sigma \sum_{l=1}^{3} b_{l,[2^k u]+1} \;G^{(l)}_{k,[2^k u]+1}(u)\right]d u\right\}. \nonumber
\end{eqnarray}
\end{widetext}
The action integral is performed by means of the quadrature scheme specified by the $4 \cdot 2^{k} = 4(n + 1)$  quadrature points 
\begin{equation}
\label{eq:ap6}
u_{ij} = 2^{-k}(\theta_i + j - 1), \quad   1 \leq i \leq 4, \ 1 \leq j \leq 2^k
 \end{equation}
and the corresponding weights 
\begin{equation}
\label{eq:ap7}
w_{ij} = 2^{-k}\omega_i.
\end{equation}
The quantities $\theta_i$ and $\omega_i$ are those from Table~\ref{Tab:I}.

The additional path variables $b_{l,j}$ make up three different layers that are additional to the layers made up by the Lie-Trotter path variables $a_{l,j}$. Such a layer $l$ is selected randomly with probability equal to the other layers. Again, due to the fact that the functions $G_{l,j}(u)$ vanish outside the interval $((j-1)2^{-k},j2^{-k})$, the variables $b_{l,j}$ from a given layer $l = 1, 2, 3$ can and must be updated independently. The appropriate weight is given by the formula 
\begin{widetext}
\begin{eqnarray}
\label{eq:ap8} \nonumber
e^{-  b_{l,j}^2/2}
 \exp\left\{-\beta \sum_{s = 1}^4 w_{sj} V\left[x_r(u_{sj})+\sigma \sum_{l=1}^{k} a_{l,[2^{l-1} u_{sj}]+1} \;{F}_{l,[2^{l-1} u_{sj}]+1}(u_{sj}) \right. \right. \\  \left. \left.+ \sigma \sum_{l=1}^{3} b_{l,[2^k u_{sj}]+1} \;G^{(l)}_{k,[2^k u_{sj}]+1}(u_{sj})\right]\right\}. 
\end{eqnarray}
\end{widetext}
For a Lie-Trotter variable $a_{l,i}$, the appropriate weight is
\begin{widetext}
\begin{eqnarray}
\label{eq:ap9} \nonumber
e^{-  a_{l,i}^2/2}
 \exp\left\{-\beta  \sum_{j = 1 + (i-1)2^{k-l+1} }^{i2^{k-l+1} }\sum_{s = 1}^4 w_{sj} V\left[x_r(u_{sj})+\sigma \sum_{l=1}^{k} a_{l,[2^{l-1} u_{sj}]+1} \;{F}_{l,[2^{l-1} u_{sj}]+1}(u_{sj}) \right. \right. \\  \left. \left.+ \sigma \sum_{l=1}^{3} b_{l,[2^k u_{sj}]+1} \;G^{(l)}_{k,[2^k u_{sj}]+1}(u_{sj}) \right]\right\}. 
\end{eqnarray}
\end{widetext}

Our choice of the $\mathrm{Ne}_{19}$ cluster for numerical experiments is motivated by the fact that the cluster presents a deep classical global minimum \cite{Doy99} that is not destroyed by the quantum effects \cite{Cal01}. In order to utilize a number of path variables that is large enough to facilitate the comparison, we conduct our computations at the low temperature of $4~\mathrm{K}$. Numerical experiments show that, at this temperature, the number of path variables that ensures a systematic error  comparable to the statistical errors is $127$ per degree of freedom (corresponding to the Trotter index $n = 31$). As opposed to the classical simulation presented in Section~II, here we do not use parallel tempering to improve the sampling. This and other techniques that are commonly used to improve the quality of the sampling have the property that they reduce the correlation of the Metropolis walker. Clearly, we do not want to measure the ability of the parallel tempering technique to do so. Therefore, we shall only conduct a simple Metropolis sampling, for each simulation. To ensure that the basin associated with the global minimum is  adequately sampled, we start all simulations from the global minimum. Had we utilized a cluster with a double funnel topology of the potential electronic surface, we could not have ensured ergodicity of the simulation at $4~\mathrm{K}$. 

Our test consists in the evaluation of the average energy and the heat capacity of the cluster using simultaneous updates of all path variables and the fast sampling algorithm, respectively.
The energy and the heat capacity estimators employed are those obtained by formal differentiation of Eq.~(\ref{eq:ap5}). The estimators have been reviewed elsewhere \cite{Pre03b, Pre03c}. In the case where all path variables are updated simultaneously, we have employed the same maximal displacements for all variables $a_{l,j}$ and $b_{l,j}$. An optimal ratio between the maximal displacements for the physical coordinate $x$ and the path variables has been determined in a separate Monte Carlo study, in which the two sets of variables were updated separately. It is however not at all clear if the ratio obtained this way remains the optimal one when all variables are updated simultaneously. This observation underlies again the better statistical efficiency of the sequential update: at the very least, the optimal displacements can be determined separately for each variable or group of variables during the original Monte Carlo simulation. Because for equilibrium properties the paths are closed (i.e., $x' = x$), the path variables from a same layer have identical marginal distributions and, therefore, they have identical maximal displacements. Thus, for the fast sampling strategy, a number of only  $1 + \log_2(n+1) + 3 = 9$ maximal displacements must be optimized. 

In both simulations, we have updated the coordinates associated with a given particle sequentially. That is, we randomly choose a particle and either update all variables (for the all-particle update strategy) or only the variables associated with a randomly chosen layer (for the fast sampling strategy). In both cases, the computational effort, as measured with respect to the number of calls to the potential, is the same. The simulations have consisted of $50$ blocks of $20$ thousand sweeps through the configuration space. Each simulation has been preceded by a number of $25$ equilibration blocks. The statistical tests described in Ref.~\onlinecite{Pre03b} have been employed to test for the independence of the block averages. 

The results of the two simulations are summarized in  Table~\ref{Tab:II}. The two sampling strategies have resulted in similar values for the average energy. However, the statistical errors are different, due to the higher correlation in the all-variable sampler. The energy statistical errors for the all-variable strategy are $2.26$ times larger than those for the fast sampling strategy. It follows that the fast sampling algorithm allows for a saving of about $80\%$ in the computational effort. For the heat capacity of the cluster, the saving is about $78\%$. For this particular example, the fast sampling algorithm has made the difference between obtaining a reliable heat capacity and not. As discussed in the preceding section, for larger numbers of path variables, larger savings in the computational effort are expected. Sure enough, the exact percentage depends not only on the quality of the sampling, but also on the smoothness of the estimator utilized. 

\begingroup
\begin{table}[!tbp]
\caption{
\label{Tab:II}
Energies, heat capacities, and associated statistical errors (twice the standard deviation) for the $\mathrm{Ne}_{13}$ cluster, with the energy expressed in units of $\epsilon_{LJ}$. }
\begin{tabular}{|c |c |c |}
\hline
Type of sampling & Energy & Heat capacity \\ 
\hline \hline
fast sampling&$-45.969 \pm 0.023$ & $0.228 \pm 0.069$\\
\hline
all variables&$-45.962 \pm 0.052$&$0.161 \pm 0.146$\\
\hline 
\end{tabular}
\end{table}
\endgroup

\section{Summary and conclusions}

The Monte Carlo sampler that uses all-particle or all-variable updates is not superior to the sequential sampler from the point of view of statistical efficiency. In fact, the sequential sampler is computationally more efficient whenever updating a single particle costs less than updating the whole potential. If one updates more than one particle at a time, the maximal displacements decrease inverse proportionally to the square root of the number of particles that are updated simultaneously. This effect has an entropic nature and appears even for independent variables. However, we warn the reader that the  statistical efficiency is only one factor controlling the rate of equilibration. The other factor is the rate at which the correlation in the generated Markov chain decays. As we have demonstrated in Section~II, save the special case mentioned above, the all-particle update and the sequential samplers share roughly the same efficiency in terms of total volume in the configuration space that is sampled for a given computational effort. Which of the two techniques have a faster equilibration time when the statistical efficiency is the same has not been decided. Nevertheless, for independent variables, the sequential update is always superior.

We have observed that, in the  L\'evy-Ciesielski form, the path variables generated by the Lie-Trotter products are grouped in a small number of layers, with the variables from the same layer being statistically independent. This property, together with the observation that a sequential sampler has better statistical efficiency for independent variables, explains the superiority of the  L\'evy-Ciesielski representation versus the other random series or normal mode approaches. In the L\'evy-Ciesielski representation, by using a sequential sampler, one can efficiently update all variables using $n\log_2(n)$ calls to the potential. For most other normal mode approaches, the scaling is $n^2$, whether a sequential or all-particle update sampler is utilized. 

To summarize, the computationally advantageous features of the L\'evy-Ciesielski implementation of path integrals are: fast computation of paths, fast path sampling, and the ability to use different numbers of path variables for the different degrees of freedom, commensurate with the quantum effects. The last property is discussed in the Appendix, where a relation between the number of time slices and the particle masses is suggested. These features  recommend the L\'evy-Ciesielski representation as a useful technique for path integral implementations.

\begin{acknowledgments} This work was supported in part by the National Science Foundation Grant Number CHE-0345280, the Director, Office of Science, Office of Basic Energy Sciences, Chemical Sciences, Geosciences, and Biosciences Division, U.S. Department of Energy under Contract Number DE AC03-65SF00098, and the U.S.-Israel Binational Science Foundation Award Number 2002170. The author wishes to thank William H. Miller and Jimmie D. Doll for helpful
discussions concerning the present work.
\end{acknowledgments}

\appendix

\section{Different numbers of time slices for different degrees of freedom}

Another advantage of the L\'evy-Ciesielski approach is that it allows for the utilization of  different numbers of time slices for  different degrees of freedom.  Li and Miller \cite{Li04} have recently shown how this must be done for general Lie-Trotter products.  For the L\'evy-Ciesielski form of the fourth order short-time approximation, the Li and Miller procedure is equivalent to utilizing a larger number of levels and quadrature points when particles with lighter masses are sampled.

 For definiteness, let us assume that the number of layers are $k$ and $k'$,  with $k > k'$. For the ``light'' coordinate, we utilize an entire random sum 
\begin{eqnarray}
\label{eq:di1} \nonumber
x_{sj} = x_r(u_{sj})+\sigma \sum_{l=1}^{k} a_{l,[2^{l-1} u_{sj}]+1} \;{F}_{l,[2^{l-1} u_{sj}]+1}(u_{sj})  \\  + \sigma \sum_{l=1}^{3} b_{l,[2^k u_{sj}]+1} \;G^{(l)}_{k,[2^k u_{sj}]+1}(u_{sj}). \quad
\end{eqnarray}
where $u_j = j / 2^k$, for $j = 0, 1, \ldots, 2^k$.
For the ``heavy'' coordinate, we utilize the independent sum
\begin{eqnarray}
\label{eq:di2} \nonumber
y_{sj} = y_r(u'_{sj})+\sigma' \sum_{l=1}^{k'} a'_{l,[2^{l-1} u'_{sj}]+1} \;{F}_{l,[2^{l-1} u'_{sj}]+1}(u'_{sj}) \\  + \sigma' \sum_{l=1}^{3} b'_{l,[2^{k'} u'_{sj}]+1} \;G^{(l)}_{k',[2^{k'} u'_{sj}]+1}(u'_{sj}), \quad
\end{eqnarray}
where 
\begin{equation}
\label{eq:di3}
u'_{ij} = 2^{-k'}(\theta_i + [j2^{k'-k}] - 1), \quad   1 \leq i \leq 4, \ 1 \leq j \leq 2^k.
 \end{equation}
For example, if $k - k' = 1$, then the consecutive values $u'_{sj}$ and $u'_{s,j+1}$ for even $j$ are equal.  When computing the second term in the sum 
\begin{eqnarray}
\label{eq:di4}
\nonumber
w_{sj} V\left(x_{sj},y_{sj}\right) + w_{s,j+1} V\left(x_{s,j+1}, y_{s,j+1}\right) \\ =
w_{sj}V\left(x_{sj},y_{sj}\right) + w_{s,j+1} V\left(x_{s,j+1}, y_{sj}\right), 
\end{eqnarray}
one exploits the fact that, at least for many empirical potentials, it is easier to compute the difference
\begin{equation}
\label{eq:di5}
V\left(x_{s,j+1}, y_{sj}\right) - V\left(x_{sj},y_{sj}\right).
\end{equation}

Assuming that the strength of the interactions felt by each of the particles is the same, the number of levels that is appropriate for each degree of freedom is determined from the condition that the physical distances spanned by the variables from the last layers be equal. As can be seen from Eqs.~(\ref{eq:di1}), (\ref{eq:di2}), and (\ref{eq:ap4}), these distances are proportional to $\sigma 2^{-k/2}$ and $\sigma' 2^{-k'/2}$, respectively. Remembering that $\sigma = (\hbar^2\beta/m_0)^{1/2}$ and $\sigma' = (\hbar^2\beta/m'_0)^{1/2}$ it follows that the numbers of layers $k$ and $k'$ must satisfy the relation
\begin{equation}
\label{eq:di6}
m_{0}2^{k} \approx m'_{0}2^{k'} 
\end{equation}
or, equivalently, the number of slices  $n  + 1 = 2^k$ and $n' + 1 = 2^{k'}$ must be in the relation
\begin{equation}
\label{eq:di7}
(n + 1) m_{0} \approx (n' + 1)  m'_{0}. 
\end{equation}
Eq.~(\ref{eq:di7}) tells us that the appropriate number of time slices is proportional to the inverse mass of the particle.

\end{document}